\documentclass[sigconf, nonacm]{acmart}

\usepackage[commandnameprefix=ifneeded,final]{changes}

\usepackage[T1]{fontenc}            
\usepackage[utf8]{inputenc}         
\usepackage{microtype}              
\usepackage[scaled=.8]{beramono}    
\usepackage[abbreviations]{foreign} 
\usepackage{hyperref}               
\usepackage{balance}                
\usepackage{cleveref}               
\usepackage{stfloats}
\usepackage{xspace}

\usepackage{soul}

\begin{document}

\title[Software Frugality in an Accelerating World: the Case of Continuous Integration]{Software Frugality in an Accelerating World: the Case of Continuous Integration}
\author{Quentin Perez}
\orcid{1234-5678-9012}
\affiliation{%
 \institution{INSA Rennes, Univ. Rennes, IRISA, Inria}
 \city{Rennes}
 \country{France}
}
\email{quentin.perez@irisa.fr}

\author{Romain Lefeuvre}
\affiliation{%
 \institution{Univ. Rennes, IRISA, Inria}
 \city{Rennes}
 \country{France}
}
\email{romain.lefeuvre@irisa.fr}

\author{Thomas Degueule}
\orcid{0000-0002-5961-7940}
\affiliation{%
  \institution{Univ. Bordeaux, Bordeaux INP, CNRS, LaBRI}
  \city{Talence}
  \country{France}
}
\email{thomas.degueule@labri.fr}

\author{Olivier Barais}
\orcid{0000-0002-4551-8562}
\affiliation{%
 \institution{Univ. Rennes, IRISA, Inria}
 \city{Rennes}
 \country{France}
}
\email{olivier.barais@irisa.fr}

\author{Benoit Combemale}
\orcid{0000-0002-7104-7848}
\affiliation{%
 \institution{Univ. Rennes, IRISA, Inria}
 \city{Rennes}
 \country{France}
}
\email{benoit.combemale@irisa.fr}

\newcommand{\github}{GitHub\xspace}
\newcommand{\githuba}{GitHub~Actions\xspace}
\newcommand{\co}{CO\textsubscript{2}\xspace}
\newcommand{\sw}{SmartWatts\xspace}

\newcommand{\secref}[1]{Section~\ref{#1}\xspace} 
\newcommand{\figref}[1]{Fi\-gu\-re~\ref{#1}}
\newcommand{\tabref}[1]{Table~\ref{#1}}

\settopmatter{printacmref=false}

\begin{abstract}
The acceleration of software development and delivery requires rigorous continuous testing and deployment of software systems, which are being deployed in increasingly diverse, complex, and dynamic environments.
In recent years, the popularization of DevOps and integrated software forges like GitLab and GitHub has largely democratized Continuous Integration (CI) \deleted{and Continuous Delivery (CD)} practices for a growing number of software.

However, this trend intersects significantly with global energy consumption concerns and the growing demand for frugality in the Information and Communication Technology (ICT) sector.
CI\deleted{/CD} pipelines typically run in data centers which contribute significantly to the environmental footprint of ICT, yet there is little information available regarding their environmental impact.

This article aims to bridge this gap by conducting the first large-scale analysis of the energy footprint of CI\deleted{/CD} pipelines implemented with \githuba \added{and to provide a first overview of the energy impact of CI.}
We collect, instrument, and reproduce 838 workflows from 396 Java repositories hosted on GitHub to measure their energy consumption.

We observe that the average unitary energy cost of a pipeline is relatively low, at 10~Wh.
However, due to repeated invocations of these pipelines in real settings, the aggregated energy consumption cost per project is high, averaging 22~kWh.
When evaluating \co emissions based on regional Wh-to-\co estimates, we observe that the average aggregated \co emissions are significant, averaging 10.5~kg.
To put this into perspective, this is akin to the emissions produced by driving approximately 100 kilometers in a typical European car (110~g\co/km).
In light of our results, we advocate that developers should have the means to better anticipate and reflect on the environmental consequences of their CI\deleted{/CD} choices when implementing DevOps practices.
\end{abstract}


\maketitle

\section{Introduction}
\label{sec:intro}

The acceleration of software development and delivery requires rigorous continuous testing and deployment of software systems, which are increasingly deployed in more diverse, complex, and dynamic environments~\cite{humble2018accelerate}.
However, this acceleration intersects significantly with concerns about global energy consumption.
The Information and Communication Technology (ICT) sector accounted for approximately 4--6\% of global energy demand in 2015, and projections suggest this could rise to 20\% by 2030~\cite{andrae2015global,malmodin2018energy,kamiya2020data}.

In recent years, integrated software forges such as GitLab and GitHub have democratized the implementation of DevOps and CI\deleted{/CD} pipelines for a growing number of software projects, compounded by the general trend towards shorter development and delivery lifecycles~\cite{humble2018accelerate}.
The modern and rich plug-in systems and associated marketplaces for CI\deleted{/CD}, such as on GitHub, make it easier than ever.
Developers can easily include pre-configured CI\deleted{/CD} actions in their projects (for quality assurance, code review, dependency management, testing, deployment, \etc) and benefit from GitHub's virtualized cloud to run their pipelines~\cite{DecanMMG22}.
GitHub even offers a free tier for its \githuba, allowing anyone to consume up to 2,000 CPU minutes per month for free.\footnote{\url{https://docs.github.com/en/billing/managing-billing-for-github-actions}}
More than ever, software can now be built, tested, and deployed at the push of a button---literally!

As a result, more than 30\% of public repositories on GitHub now have at least one CI\deleted{/CD} workflow implemented~\cite{bouzenia2023resource}.
A workflow consists of a set of jobs tied to repository events (\eg a push, a new pull request, a release) or running periodically, which performs a sequential list of steps when triggered.
Given the sheer number of repositories on GitHub (420+ million at the time of writing)\footnote{\url{https://github.com/about}} and the potentially workflow-triggering activity on these repositories, we hypothesize that CI\deleted{/CD} activity on GitHub may incur a significant cost in terms of energy consumption.
The accessibility and ease of use of modern CI\deleted{/CD} platforms exacerbate this issue.

Several authors have analyzed CI\deleted{/CD} pipelines on GitHub, for instance to study the motivation of developers for using such tools~\cite{kinsman2021software}, the types and frequency of events that trigger workflows~\cite{DecanMMG22}, and the monetary costs of CI\deleted{/CD}~\cite{bouzenia2023resource}.
However, there is little information available regarding the overall energy consumption of CI\deleted{/CD} pipelines.
Studies on developer practices have shown that the energy usage of applications is a common concern, though it varies with the application domain.
For example, Manotas \etal observed that nearly 60\% of interviewed practitioners reported that energy usage could influence their future developments~\cite{manotas_empirical_2016}.
40\% of the practitioners stated that they occasionally work on projects with energy usage requirements.
This proportion drops to 27\% when considering responses from developers of applications running in data centers, where the energy usage of their code is not a first-class requirement, unlike applications running on battery-powered devices like smartphones.
This study also highlights that addressing such concerns is hindered by a lack of accessible information and techniques for designing energy-efficient systems and for estimating the energy consumption of code.


In this article, our goal is to fill this gap by conducting the first large-scale analysis of the energy consumption of GitHub \deleted{CI/CD} workflows.
We create a controlled and reproducible environment by executing workflows locally on a regular server to measure their energy consumption.
\added{We focus on CI workflows that can be executed in our experimental environment and elude Continuous Delivery (CD) pipelines that require access to external infrastructure and might partially fail.}
We execute 838 workflows from 396 popular Java repositories\added{, using the number of stars as a proxy for popularity~\cite{BorgesHV16}.} \added{Then, we measure} the energy consumption of their CI workflows by instrumenting our measurement machine with \sw, a software power meter included in the PowerAPI toolbox.

We find that the mean energy consumption of individual workflow runs is relatively low at 10.2~Wh (similar to the energy consumption of a 10.5~W LED light bulb for an hour), with 20 workflows consuming more than 100~Wh, up to 703~Wh.
When multiplying the energy consumption of individual runs by the number of times they were triggered, however, we find that their aggregated consumption reaches a mean of 22~kWh, similar to the consumption of a 1150~W microwave for 3 minutes per day over a year.

Implementing CI\deleted{/CD} pipelines has become so commonplace that users may not always carefully consider the impact of their decisions.
In light of our results, we advocate for providing users with the means to estimate the impact of their choices on energy consumption.
We believe that the necessary growth of DevOps and CI\deleted{/CD} can go hand in hand with more sustainable software practices, and we hope the community will take this matter on fully.

\section{Continuous Integration Systems and their usages}
\label{sec:cicd_usages}

Our world is accelerating, and software is not exempt from this acceleration~\cite{humble2018accelerate}. Thus, as software becomes more prevalent in our lives, software systems shape the world around us. Mastering the quality of the produced software becomes essential in the face of its omnipresence. More and more, companies are turning to DevOps approaches to control quality and maintain a rapid delivery pace~\cite{wakatareKKSHIK19,HiltonTHMD16}. DevOps is both a set of methods for software development and maintenance and an accompanying set of tools and technologies~\cite{ErichAD17}. It encompasses a lifecycle that spans both development phases (design, implementation, construction, testing, \etc) and more operational phases (deployment, monitoring, \etc).

DevOps practices involve using a variety of methods and tools to cover different stages of this lifecycle, including implementation, testing, construction, deployment, and monitoring. Automated software build managers such as Gradle, Maven, NPM, Cargo, and CMake are used to manage the development part of the DevOps cycle. They often integrate dependency management and plugins for software construction. They can be configured according to different needs, such as the use of caches for dependencies, language versions for compilation, or the executable format generated. This configurability is enhanced by the use of plugins, allowing, for example, script execution or test automation. The execution of these build managers is often carried out by a CI\deleted{/CD} system executed on a local development laptop or in a virtualized cloud environment. These systems enable Continuous Integration (CI) and Continuous Delivery (CD) and are pillars of the DevOps practice. They address the need for agility and accelerated development while maintaining quality.

The principles of continuous integration have proven their value in maintaining and improving software quality~\cite{VasilescuYWDF15,HiltonTHMD16}. Initially, CI\deleted{/CD} platforms were software systems physically installed on local CI\deleted{/CD} machines. However, the rise of online platforms such as Circle CI and Travis CI and software forges such as GitHub and GitLab has largely democratized the use of CI\deleted{/CD}. This is notably due to their ease of setup and the availability of virtualized execution environments in the cloud. CI\deleted{/CD} systems perform a variety of tasks, including cloning software repositories, executing build managers, generating documentation, conducting static and dynamic analyses, and building and running containers. These tasks are materialized on the CI\deleted{/CD} platform as a process called a pipeline. Each pipeline can then be executed on demand. Pipelines are often run periodically or triggered by project-specific events (such as commits or pull requests on software forges)~\cite{DecanMMG22}. Specific platforms define their own events. For example, \github allows running CI\deleted{/CD} pipelines when opening an issue or preparing a release.
Practically speaking, pipelines are most often defined through descriptors written in a particular domain-specific language (DSL). Each CI\deleted{/CD} system defines its own DSL with a clear operational semantic to capture the different processes involved, such as build \deleted{, deployment,} or configuration.
 
The creation of pipelines is not bound by a specific design pattern, but best practices have emerged in the scientific and gray literature~\cite{ZampettiVPCGP20,VassalloPJGP20,VassalloPGP19}. These practices address requirements such as maintainability, usability, performance, \etc. While their technical aspects are not in question, the issue of energy consumption is raised. This is particularly relevant as platforms like Travis CI or \github offer free services that allow the creation and execution of one or more pipelines for a software project with minimal restrictions.

\begin{figure}[t]
\centering
\includegraphics[width=\linewidth]{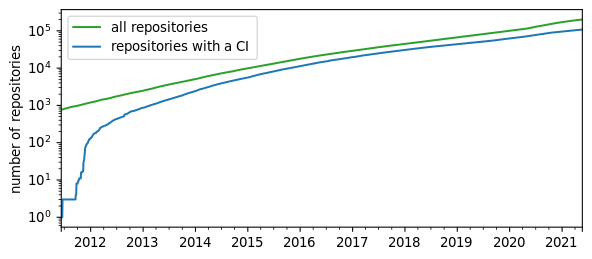}
\caption{Evolution of the number of repositories (green line) and number of repositories using a CI (blue line), from Golzadeh \etal \cite{GolzadehDM22}.}
\label{fig:repo_with_ci}
\end{figure}

Golzadeh \etal studied the use of CI\deleted{/CD} systems in \github repositories~\cite{GolzadehDM22}. \Cref{fig:repo_with_ci} illustrates the evolution of the popularity of CI over time in Golzadeh \etal's dataset. The number of CI systems tends to follow the number of repositories. Starting June 2011, there has been exponential growth in the number of CI systems, mainly due to the introduction of Travis CI and its integration with GitHub. Today, numerous actors like Travis CI, \githuba, Microsoft Azure, Circle CI, and others have invested in the CI market. According to Golzadeh \etal, by the end of their study in 2021, around 57.3\% of repositories were using one or more CI systems, with a mean number of $1.59$ pipeline per repository. This staggering growth in the number of CI\deleted{/CD} pipelines raises concerns about the environmental impact of modern CI\deleted{/CD} practices in general, and on \github in particular. Each pipeline incurs a unitary cost, which is then multiplied by its execution frequency. As the execution frequency scales with the frequency of events in a repository, the modern practice of accelerating software development \deleted{and deployment} raises serious questions regarding its environmental impact.
\added{Bouzenia and Pradel showed that it is possible to estimate the cost of pipelines from their execution time. To provide an estimation of the costs, they studied the amount of resources used for pipeline execution in terms of \githuba virtual machines and execution time. They proposed strategies to be more economically efficient, such as failing fast, caching, and skipping workflows~\cite{BouzeniaP24}.}

\added{A recent exploratory study provides first insights on the environmental impact of CI systems. Zaidman evaluates the energy consumption of machines running the CI pipelines of 10 open-source Java projects and shows that their energy footprint is non-negligible when considering the cumulated energy consumed over a year~\cite{Zaidman24}.}

\added{Building on these early results, we pursue the objective of exploring the energy footprint of CI systems with two main novelties. We conduct the first large-scale study of the pipelines of 396 popular repositories hosted on \github. We use a software power meter in a controlled environment to isolate and estimate the energy consumption induced by pipeline executions rather than measuring the total energy consumption of the machine to obtain more accurate results.}

\section{Measuring Energy Consumption of Software}
\label{sec:measuring_energy_software}

One of the challenges in estimating software energy consumption is energy measurement. Software can utilize multiple hardware resources simultaneously (CPU, memory, HDD, network, etc.), and the architectures of CPUs and GPUs are becoming increasingly complex. Operating Systems (OS) are also complex software systems, not only due to hardware complexity but also due to requirements for performance, security, usability, \etc. One way to measure energy consumption is by monitoring hardware with power meters on power distribution units or directly on the power supply. However, in the case of CI\deleted{/CD} pipelines, it is not possible to have a precise view of which process is being executed. Faced with the lack of dedicated measuring tools for software, Intel created the Running Average Power Limits (RAPL) in 2011 to monitor the energy consumption of Intel CPUs~\cite{DavidGHKL10}. RAPL is a set of counters directly integrated into the CPU. The RAPL sampling rate is high, with each energy value measured every 1 ms (1000~Hz)~\cite{JayOLTOF23}. RAPL provides information about energy consumption and allows restricting power usage across various levels or power domains: the entire CPU socket, all CPU cores, integrated graphics processing unit, dynamic random-access memory and the entire System on Chip. Each power domain is accessible from the operating system through a Model-Specific Register. In 2017, AMD implemented a similar system in their CPUs with Zen architecture. Although RAPL provides a solution focusing on energy consumption at the CPU level, it is difficult to isolate the power consumption of a specific process or program. 

To overcome the limitations of RAPL, Fieni \etal~\cite{FieniRS20} created \sw in 2019. \sw is included in the PowerAPI toolkit,\footnote{https://powerapi.org/} which allows the deployment of software power meters.
\sw is a software power meter that allows the isolation of energy consumption in Docker containers and processes through Control Groups (CGroups) \cite{Rub1}. Control groups were created by two Google engineers, Paul Menage and Rohit Seth, to limit, monitor, and isolate processes running on Linux servers. For example, CGroups can limit the CPU and memory resources of a specific group of containers. A specific sensor (HWPC-Sensor) included in PowerAPI is used to collect RAPL and performance values (CPU usage, last-level cache miss, \etc) coming from the RAPL interface and CGroups. After that, the \sw power meter uses a regression based on these values to calculate power in watts at a time $t$.

\sw offers a reliable sampling frequency of 2~Hz~\cite{FieniRS20}. It allows storing data in various formats and databases, including JSON, CSV, Mongo, InfluxDB, etc. Another power meter solution is Scaphandre, which emerged at the end of 2020. Scaphandre uses CPU usage and RAPL to estimate the power at a specific time. Scaphandre also provides storage features in various formats: JSON, Prometheus, Riemann, \etc However, the maximum sampling frequency of Scaphandre is 0.5~Hz~\cite{JayOLTOF23}, which is lower than that of \sw.

\section{Studying the Energy Consumption of Pipelines}

As discussed in \secref{sec:measuring_energy_software}, there are various tools available for monitoring and storing energy consumption values. However, monitoring the energy consumption of pipelines has some unique characteristics. Pipelines use a dedicated program called a runner to execute the pipeline. The runner is responsible for reading the pipeline descriptor file and executing its steps. Each step may utilize one or more tools. Therefore, monitoring energy consumption in pipelines requires monitoring both the runner and all the processes it launches. Another specificity of the pipeline mechanism is the execution of containers for building steps. Consequently, monitoring tools need to be capable of capturing the energy consumption of all subprocesses and containers launched by the runner. 

\added{Our methodology is based on CPU energy measurement as an indicator of the impact of a CI run.  We use software-based solutions to assess energy consumption, with the aim of making our methodology easily reusable in industrial contexts that require non-invasive measurement.}
\added{We focus on CPU energy consumption for practical reasons, as \sw does not support assessing RAM energy consumption on all devices, including our own. Additionally, we are not aware of software tools available for evaluating or estimating disk energy consumption. Fine-grained measurements of disk consumption require hardware sensors but this method is unable to distinguish the consumption linked to a specific pipeline subprocess.}
\added{Our choice is also motivated by previous studies, which have shown that CPU consumption tends to be the main source of energy consumption in servers. For example, Malladi \etal~\cite{malladiNPLKH12} found that in their context, with similar hardware, CPU energy accounted for 60-69\% and RAM accounted for 6-19\% of the total energy consumption. Finally, quantifying network energy costs is extremely challenging as it requires knowledge of network infrastructure and geographical boundaries~\cite{guennebaud24}.}

\sw \cite{FieniRS20} is the best candidate for this use case. In this experiment, we study pipelines, also known as workflows, from \github. \github has its own service for running workflows in the cloud: \githuba. By default, \githuba uses a runner in the cloud, but it provides the option to execute the runner locally on a dedicated machine. This enables the monitoring of the runner's energy consumption. 

\begin{figure*}[t]
  \centering
 \includegraphics[width=0.8\linewidth]{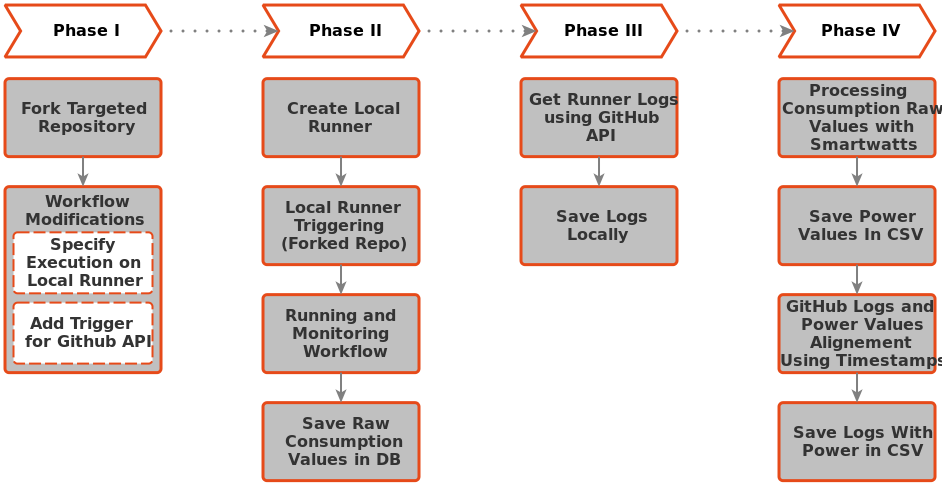}
  \caption{\label{fig:workflow_monitoring_global}Process for extracting, measuring and saving the consumption of dataset pipelines.}
\end{figure*}

\figref{fig:workflow_monitoring_global} provides an overview of the process for extracting energy consumption using \github, \githuba runner, and \sw.
The first phase entails forking the targeted GitHub project, a crucial step to obtain the rights to modify and execute workflow files.
We modify the workflow files by introducing the \textit{repository\_dispatch} trigger, enabling workflow activation via \github's API and directing execution to our local runner machine rather than GitHub's cloud.
These modifications allow us to automatically run and monitor the pipelines of our dataset.

The second phase includes triggering the workflow using the \github API to run it locally on a dedicated measurement machine. Our measurement machine, a Dell Precision Tower 5810, is equipped with an Intel Xeon E5-1603v4 at 2.8~GHz, 8~GB of DDR4 at 2400~MHz, and Xubuntu 22.04.2 LTS with kernel version 6.2.0-35-generic. This machine is equipped with a specific software sensor from PowerAPI that collects metrics used by \sw. The local runner is added to a Control Group, enabling the capture of metrics by the sensor. The runner is added to the Control Group with a ``sticky'' option, ensuring that each subprocess launched by the runner stays within the Control Group. Each pipeline is executed sequentially to maintain consistency in measurements. Since each measurement is time-stamped, it would be impossible to maintain temporal consistency with a parallel execution. This setup allows for the collection of metrics specific to the GitHub runner and associated tools. The sensor collects values at a frequency of 2~Hz (500 ms)~\cite{FieniRS20}, and all values are stored in a MongoDB database. 

The third phase involves deleting the runner created for the pipeline and retrieving formatted logs from \githuba using the \github API. This step prevents the creation of multiple local runners and avoids confusion during power measurement.

In the final phase, \sw computes the power values from the raw metrics collected by the sensor. The result is a CSV file with timestamps and associated power values. By using the timestamped log files and the CSV file, it becomes possible to map the power values to the corresponding logs. This alignment is necessary to identify the energy measurements corresponding to the start and stop times of the pipeline. The pipeline log is timestamped with the beginning time and the end time, while measurement values are timestamped with a sampling rate of 500~ms. The alignment phase consists of identifying measurement values corresponding to the duration of the pipeline execution using timestamps. Finally, we obtain a file containing pipeline extraction logs and the corresponding energy consumption values.

To build our dataset, we extract \github projects using its API. We select repositories written in Java with a minimum of 100 stars that have at least one GitHub Actions workflow and utilize a build manager (Gradle or Maven). The latter ensures that these projects are more likely to include a DevOps pipeline with build phases that go beyond documentation assembling or website building. We obtain an initial set of 1,500 Java projects complying with these criteria. Finally, we attempt to execute the workflows of each repository and filter out those that cannot be executed and those that do not expose the execution history of the workflows. Our final dataset comprises 396 repositories.

Among these 396 repositories, 838 \github \deleted{CI/CD} pipelines were executed once locally on our measurement machine, resulting in an average of 2.12 pipelines per project. \added{The average execution time for pipelines in our monitored environment is 143 seconds. Given this average, the 500~ms sampling rate of our methodology is acceptable. As we mined popular Java projects, the majority of our pipelines are used jointly with build managers such as Maven or Gradle.} 
\tabref{tab:describe_dataset} gives some descriptive statistics of the studied repositories.
All of these 838 pipelines executed on the \githuba platform have a history of executions (runs). For each pipeline, all the histories of runs on \githuba platform have been retrieved using \github API.
Thus, we have collected 293,281 runs of pipelines with their associated execution triggers.


\begin{table}[]
    \centering
        \begin{tabular}{l|rr}
        \toprule
           & \#Java LoC & \#Java Files \\
            \midrule
                mean & 802,60.1 & 774.8 \\
                std & 238,943.8 & 2,129.9 \\
                min & 42.0 & 3.0 \\
                25\% & 8,592.2 & 106.0 \\
                50\% & 23,352.5 & 290.0 \\
                75\% & 65,695.8 & 703.5 \\
                max & 3,854,187.0 & 33,471.0 \\
            \bottomrule
        \end{tabular}
    \caption{Statistical description of the 396 \github projects in the dataset}
    \label{tab:describe_dataset}
\end{table}

\section{The environmental footprint of CI pipelines}

\begin{figure*}[tb]
  \centering
  \includegraphics[width=\linewidth]{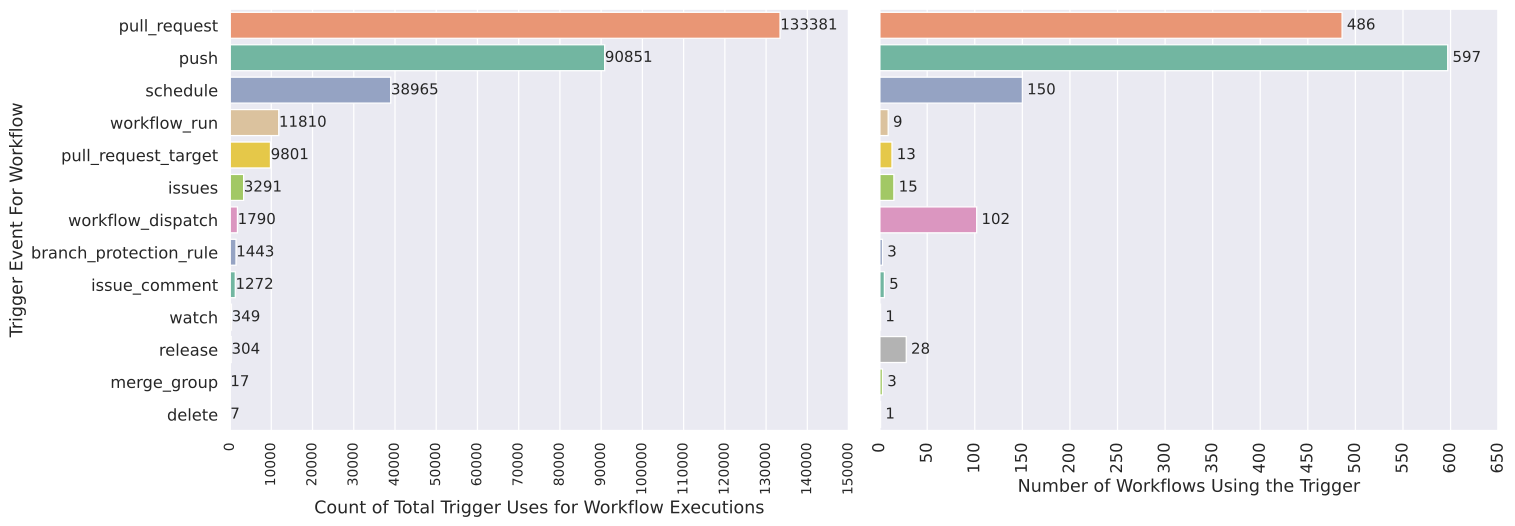}
  \caption{\label{fig:countplot_trigger}Count of each trigger type used in our dataset.}
\end{figure*}

As shown by Golzadeh \etal \cite{GolzadehDM22} and described in \secref{sec:cicd_usages}, CI\deleted{/CD} platforms are widely used today in \github repositories. Observations conducted on our dataset of 396 projects with a history of executions on \github reveal 293,281 executions over 838 pipelines. 

An interesting aspect of pipeline execution is the variety of triggers, which essentially act as launchers for the pipeline. These triggers can be integrated into a single pipeline, allowing multiple triggers to initiate the same pipeline.
\figref{fig:countplot_trigger} presents the different kinds of triggers used in our dataset of 396 \github projects with a history of executions. The most used trigger is on pull requests. Pull requests are contributions submitted by developers to repositories on \github and subsequently reviewed and integrated by the maintainers or owners. This trigger is interesting because, upon integration, project maintainers aim to ensure the absence of regressions.

The second most used trigger is the push event. The use of this trigger can be questioned due to the potential high frequency of commits and the nature of changes in the repository. While many changes may involve code modifications requiring a pipeline execution to prevent regression, running the complete pipeline, including compilation and testing phases, may seem excessive in cases where the change is unlikely to require a full CI\deleted{/CD} pass, such as when modifying documentation or auxiliary files.

The third most used trigger concerns scheduled builds. \github developers using \githuba can schedule builds periodically (every hour or day, once per week, \etc). Scheduled triggers guarantee a build independently of any modifications to the repository but can still be useful to check, for instance, that changes external to the project (such as new dependency releases) do not affect its functionalities.

\begin{figure}[tb]
  \centering
 \includegraphics[width=\linewidth]{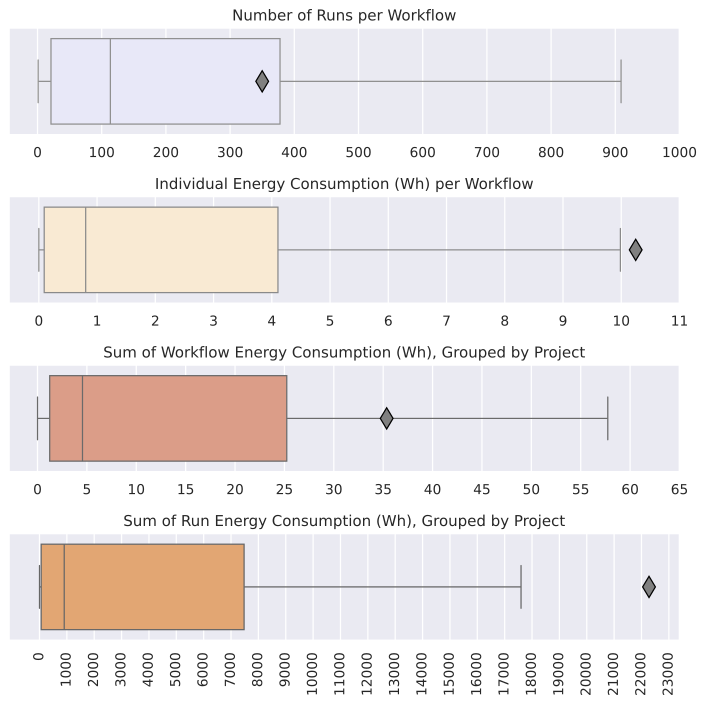}
  \caption{\label{fig:boxplot_energy_worlflows_runs}Boxplots showing the distribution of the number of runs and energy consumption for workflows and runs in our dataset of 396 \github projects.}
\end{figure}

\figref{fig:boxplot_energy_worlflows_runs} and \tabref{tab:describe_energy_worlflows_runs} present the number of runs in projects and the individual energy consumption of workflows. They also illustrate the overall energy consumption of workflows and the consumption attributed to pipeline executions (runs), grouped by project. The individual energy cost is, on average, relatively low, with a median of 0.8~Wh and a mean of 10.2~Wh. The latter is equivalent to the energy consumption of a 10.5 W LED light bulb\footnote{Light Bulb Datasheet: \url{https://www.lighting.philips.co.uk/consumer/p/led-bulb/8718699763275/specifications}} for approximately an hour. While the individual consumption may seem marginal, the cumulative cost multiplies with the number of executions. Moreover, we identified 20 pipelines consuming more than 100~Wh, with a maximum consumption of 703~Wh.

\begin{table*}[tb]
    \begin{tabular}{lrrrr}
      & \#Run     & \begin{tabular}[c]{@{}r@{}}Individual Workflow Energy\\ (Wh)\end{tabular} & \begin{tabular}[c]{@{}r@{}}Sum of Workflow Energy Consumption, \\ Grouped by Project (Wh)\end{tabular} & \begin{tabular}[c]{@{}r@{}}Sum of Run Energy Consumption\\ Grouped by Project (Wh)\end{tabular} \\
    \midrule
        mean & 349.9773 & 10.2479 & 35.3527 & 22275.4134 \\
        std & 741.6123 & 41.4801 & 78.2955 & 55686.2248 \\
        min & 1.0000 & 0.0003 & 0.0020 & 0.0246 \\
        25\% & 21.0000 & 0.0923 & 1.2355 & 62.6426 \\
        50\% & 113.5000 & 0.8045 & 4.5554 & 904.0360 \\
        75\% & 377.7500 & 4.1058 & 25.2379 & 7477.5850 \\
        max & 10674.0000 & 703.3086 & 777.4863 & 290381.1909 \\
    \end{tabular}
    \caption{\label{tab:describe_energy_worlflows_runs}Statistical description of the energy consumption for workflows and runs.}
\end{table*}

The number of runs plays a crucial role in determining the final energy consumption. Our results indicate that the mean number of runs per workflow is not negligible in the case of popular Java projects. We observe a mean of 350 runs and a median of 113. To calculate the total energy consumption of pipelines by project, we multiply the number of runs for each workflow by their respective energy consumption. This gives us a global estimate of the consumption of pipelines over all their runs. Naturally, the workflow and the underlying project may have evolved over time so this estimate is not perfectly accurate, but it gives us a good idea of the energy implications of multiple runs.
The energy consumption of pipelines found is significantly high, with a median consumption of 861.6~Wh per project and a mean consumption of 22,275.4~Wh. The mean is equivalent to the consumption of a 1150~W microwave\footnote{Microwave Datasheet: \url{https://downloadcenter.samsung.com/content/UM/202209/20220909191408420/User_MS23F301TAK_EU_DE68_04182K_07_EN.pdf\#page=28}} used 3 minutes per day during one year (20,987~Wh). 

\begin{figure*}[tb]
  \centering
  \includegraphics[width=.9\linewidth]{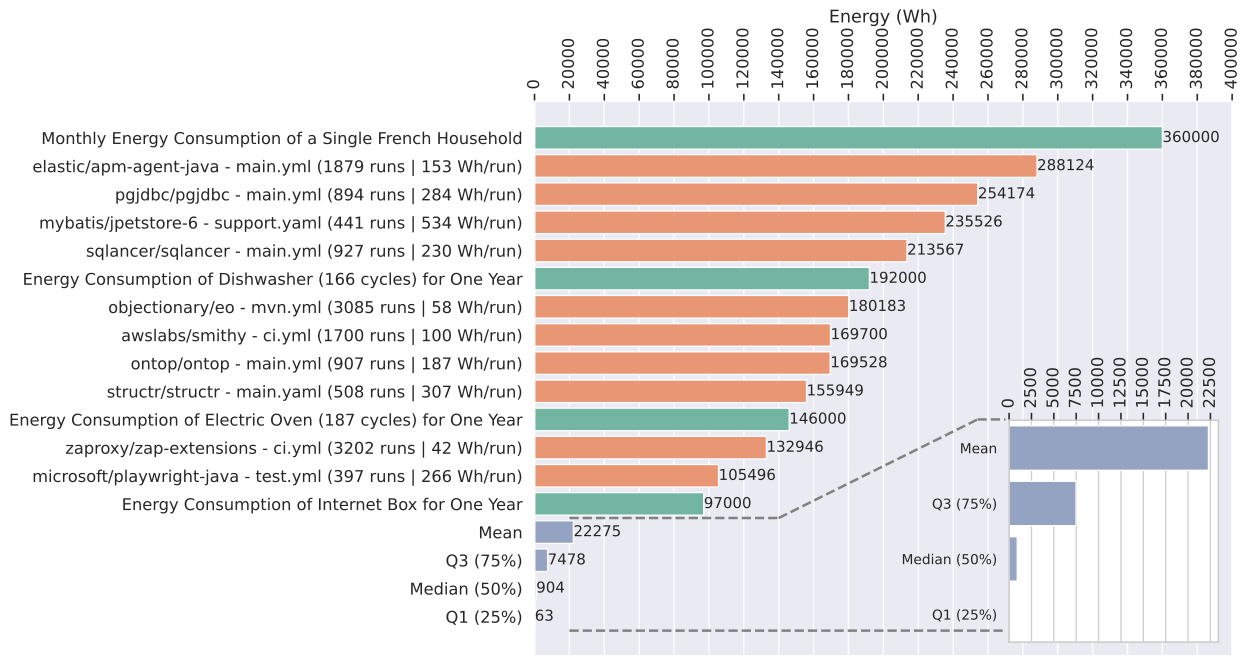}
  \caption{\label{fig:barplot_10_most_consumers}Ten most energy-hungry workflows (individual workflow energy multiplied by the number of runs) compared with four reference assets and statistical data.}
\end{figure*}

\figref{fig:boxplot_energy_worlflows_runs} does not show outliers. However, we notice some highly energy-hungry pipelines. To provide a more global view of pipelines with high energy consumption, we present \figref{fig:barplot_10_most_consumers}. \figref{fig:barplot_10_most_consumers} shows the ten most energy-consuming pipelines. We provide different reference points~\cite{ademeusages2021} for comparison using four different assets: a French household, a dishwasher, an electric oven, and an internet box. Four pipelines consume more than a dishwasher over one year. Of the ten pipelines, eight have a high individual energy consumption:~they consume more than 100~Wh for each execution. This is 10 times the mean of our dataset. Only two pipelines consume less than 100~Wh for their individual execution.

We can also view energy consumption from another perspective, which is that of \co emissions associated with pipelines. The quantification in terms of energy consumption can be converted to \co emissions using the mean of emissions in grams per Wh according to different areas in the world. Hence, we collected the mean of 4 geographical areas:
\begin{itemize}
    \item France: 68 g\co/kWh (2023) \cite{emissionsEEA}
    \item Europe: 251 g\co/kWh (2023) \cite{emissionsEEA}
    \item USA: 190 g\co/kWh (2023) \cite{emissionsUSA}
    \item World: 475 g\co/kWh (2019) \cite{emissionsWorld}
\end{itemize}

\begin{figure*}[tb]
  \centering
  \includegraphics[width=.9\linewidth]{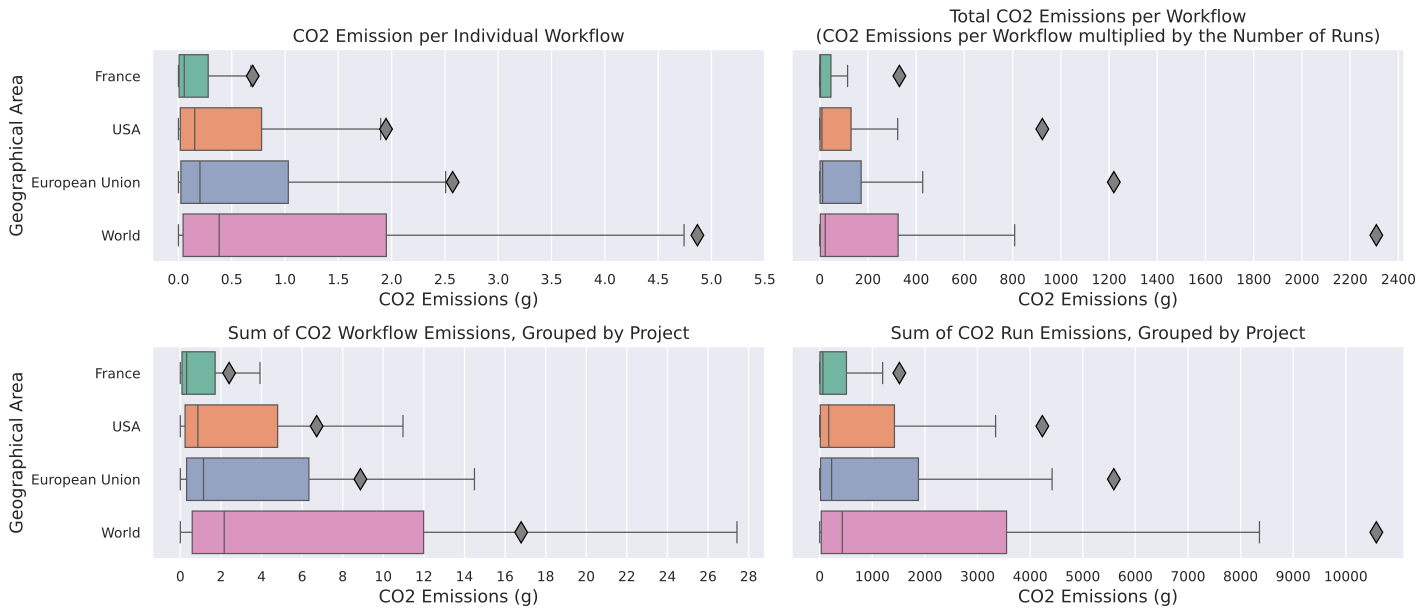}
  \caption{\label{fig:boxplot_co2}Boxplots showing the distribution of the number of runs and \co emissions for workflows and runs in our dataset considering four different regions.}
\end{figure*}

\begin{figure}[tb]
  \centering
  \includegraphics[width=1\linewidth]{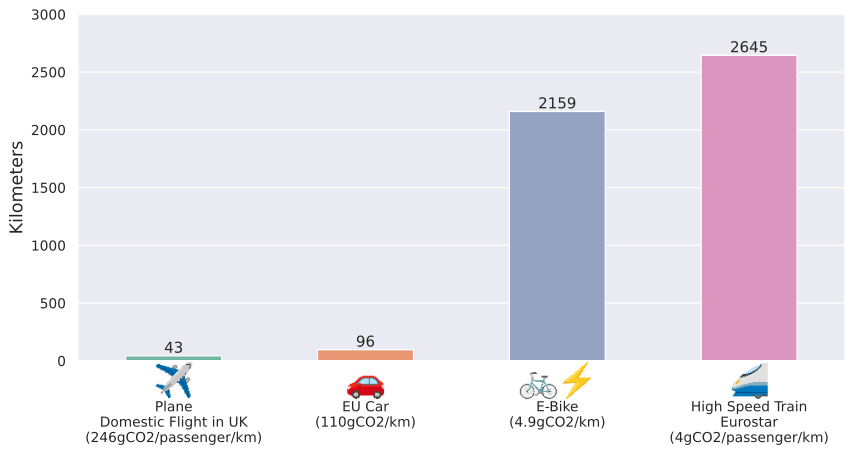}
  \caption{\label{fig:barplot_co2_km}Bar chart representing the kilometers that can be traveled according to 4 modes of transportation, considering the average world \co emissions (10.5~kg) for all runs grouped by project.}
\end{figure}

As done in \figref{fig:boxplot_energy_worlflows_runs}, in \figref{fig:boxplot_co2} we calculate \co emissions for the individual workflow, the overall workflow emissions (individual workflow emissions multiplied by the number of runs), the sum of individual workflow emissions grouped by project, and the total emissions for all runs across all workflows grouped by project. We observe the same thing with \co emissions than with energy consumption. The unit cost in terms of emissions is low, but it is highly dependent on the geographical region and its typical energy mix. The nuclear part in the energy mix is important in France (68\%)~\cite{nuclearmixFrance2020} and a fifth in the USA (19\%) \cite{nuclearmixUSA2020}, which explains the mean values for unit consumption: 0.75 and 1.9 grams of \co. However, if we take the mean of \co emission per kilowatt-hour in the world, electricity being less decarbonized, we obtain values 6 times higher than in France and 2.5 times higher than in the United States. The cost of emissions for all runs per project is relatively high if we consider average global emissions. 
We obtain an average of 10.5~kg of emitted \co. \figref{fig:barplot_co2_km} illustrates the translation of this average into kilometers traveled using 4 modes of transportation:
\begin{itemize}
    \item Plane (domestic flight in UK): 246~g\co/passenger/km \cite{owid-travel-carbon-footprint}
    \item Car in Europe: 110~g\co/km \cite{emissionscarsEU2022}
    \item E-Bike: 4.9~g\co/km \cite{mcqueen2020bike}
    \item High speed train (Eurostar): 4~g\co/passenger/km \cite{owid-travel-carbon-footprint}
\end{itemize}
These 10.5~kg of \co are equivalent to flying 43~km on a domestic flight in the UK, driving \added{96}~km with a car at the European average emission level, cycling 2159~km with an electric bike, or covering 2645~km on the Eurostar.

Although individually low in energy consumption, workflows consume a lot of energy when executed at a high frequency. Especially, pipeline triggers on commits (push events) are among the most frequently used.
Regional variance plays a crucial role in the \co emissions. Emissions related to a workflow can be multiplied by a factor of 6 between two regions. This issue of deployment in a specific region is not unique to CI\deleted{/CD} systems but inherent in IT systems. Current CI\deleted{/CD} systems executed in the Cloud like \githuba do not allow deployment in a particular geographical area, as is the case with Amazon Web Services, for example.

\section{Threats to Validity}
\label{sec:ttv}

In this paper, we estimate the energy consumption of CI\deleted{/CD} pipelines by reproducing them on a single end-user machine.
This approach offers several advantages.
First, it allows for a more controlled and reproducible environment, minimizing external variables that could impact the measurements.
Second, it ensures that we can implement a precise measurement methodology using the appropriate tools and methods, as described in~\Cref{sec:measuring_energy_software}.
Finally, it makes the measurements more relatable for most developers who may not have expert knowledge of the internals of data centers but may have already administered a CI\deleted{/CD} toolchain on a personal machine or server and have a sense of the footprint of individual households.

However, there are also \replaced{limitations}{drawbacks} to this methodology. \added{As we analyze Java-based \githuba pipelines and conduct our experimentation in a controlled environment, our findings cannot be generalized to other execution environments or projects written in other programming languages.} Data centers often use hardware optimized for running such intensive pipelines and are known to be more energy-efficient, at scale, than regular machines~\cite{prieto2022energy}.
Indeed, it is common to execute pipelines using preemptible instances on underutilized cloud servers, such as AWS Spot Instances.
Power and energy management within data centers cannot be directly compared to an end-user machine, so we do not claim that our results are representative of the energy that these pipelines would consume in real-world settings.
Nonetheless, our methodology has the merit of highlighting several issues with the current implementation of CI\deleted{/CD} in software forges, and we hope that future research will supplement our findings with on-site experiments.

The sampling rate we use for measurement is fixed at 2~Hz (500~ms) as recommended by Fieni \etal~\cite{FieniRS20}.
However, after our experiments, Jay \etal showed that the sampling rate was reliable up to 10~Hz~\cite{JayOLTOF23}, and we plan to refine our measurements in the future.
\added{Some CI workflows require external credentials to interact with access-controlled resources or infrastructures, for instance to publish an artifact publicly. The energy consumption of that interaction is not handled with our methodology.}
Finally, our methodology focused on a medium-sized sample of popular Java projects hosted on GitHub and using GitHub Actions.
Analyzing software projects hosted elsewhere or written in other programming languages and using different build systems and workflow steps may yield significantly different results.

This paper is a preliminary step towards gaining a better understanding of the environmental impact of modern DevOps practices.
We hope that our methodology will encourage researchers and practitioners to expand it beyond our dataset to encompass various programming languages and environments. 


\section{Conclusion and Challenges}

DevOps is becoming increasingly widespread, and so are the tools and practices that accompany it. In this context, the ease with which pipelines can be created on CI\deleted{/CD} platforms and integrated with repositories such as on \github has led to their rapid growth. This growth is not without impact, even if it may seem transparent to the user. Using the cloud abstracts the energy dimension, but it should not be forgotten that CI\deleted{/CD} pipelines run on physical machines. While the energy consumption of individual pipelines may not be that high, the multiplicity of triggers and executions results in substantial energy consumption.

Reducing energy consumption can be achieved by refining the triggers used and excluding pipelines from execution when certain files, such as documentation and auxiliary files, are modified. Another important thing is the feedback given about energy consumption about the pipeline. However, there are no bespoke tools integrated into CI\deleted{/CD} systems to measure and report on the energy consumption of pipelines. Developers cannot control their impact without timely and relevant feedback during the development process. This feedback can be within the CI\deleted{/CD} system, but it can just as well be in the IDE. It could be possible to categorize pipelines according to different patterns or practices to define a catalog of best practices to avoid excessive energy consumption. However, today, this set of practices and tools does not exist for developers and project decision-makers despite its huge potential impact. One explanation lies in the fact that energy consumption in software engineering is an emerging field, and all the more so in the case of DevOps, which is part of the software lifecycle. The immateriality of software, unlike a physical product, makes it difficult to apprehend energy consumption in the lifecycle. This paper sheds some first light on this issue and highlights the importance of working on the tools gravitating around the software product and decision-makers. This importance is reinforced by the urgency of global warming and a context of uncertainty about the future, electricity production methods, and the amount of energy available.

\bibliographystyle{acm}
\bibliography{main}

\begin{thebibliography}{10}

\bibitem{emissionsUSA}
{\sc Administration, U. E.~I.}
\newblock How much carbon dioxide is produced per kilowatthour of u.s. electricity generation?, 2023.
\newblock Accessed: 04-12-2023.

\bibitem{emissionscarsEU2022}
{\sc Agency, E.~E.}
\newblock Average emissions from new cars and vans in europe continue to fall, according to provisional data, 2022.
\newblock Accessed: 04-12-2023.

\bibitem{emissionsEEA}
{\sc Agency, E.~E.}
\newblock Greenhouse gas emission intensity of electricity generation, 2023.
\newblock Accessed: 04-12-2023.

\bibitem{andrae2015global}
{\sc Andrae, A.~S., and Edler, T.}
\newblock On global electricity usage of communication technology: trends to 2030.
\newblock {\em Challenges 6}, 1 (2015), 117--157.

\bibitem{nuclearmixUSA2020}
{\sc Association, W.~N.}
\newblock Nuclear power in the usa, 2020.
\newblock Accessed: 07-12-2023.

\bibitem{nuclearmixFrance2020}
{\sc Association, W.~N.}
\newblock Nuclear power in france, 2021.
\newblock Accessed: 07-12-2023.

\bibitem{BorgesHV16}
{\sc Borges, H., Hora, A.~C., and Valente, M.~T.}
\newblock Understanding the factors that impact the popularity of github repositories.
\newblock In {\em 2016 {IEEE} International Conference on Software Maintenance and Evolution, {ICSME} 2016, Raleigh, NC, USA, October 2-7, 2016\/} (2016), {IEEE} Computer Society, pp.~334--344.

\bibitem{bouzenia2023resource}
{\sc Bouzenia, I., and Pradel, M.}
\newblock Resource usage and optimization opportunities in workflows of github actions.
\newblock In {\em 46th {IEEE/ACM} International Conference on Software Engineering ({ICSE})\/} (Lisbon, Portugal, April 2024), {ACM}, pp.~25:1--25:12.

\bibitem{BouzeniaP24}
{\sc Bouzenia, I., and Pradel, M.}
\newblock Resource usage and optimization opportunities in workflows of github actions.
\newblock In {\em 46th {IEEE/ACM} International Conference on Software Engineering ({ICSE})\/} (Lisbon, Portugal, April 2024), {ACM}, pp.~25:1--25:12.

\bibitem{DavidGHKL10}
{\sc David, H., Gorbatov, E., Hanebutte, U.~R., Khanna, R., and Le, C.}
\newblock {RAPL:} memory power estimation and capping.
\newblock In {\em 16th ACM/IEEE International Symposium on Low Power Electronics and Design\/} (Austin, Texas, USA, August 2010), {ACM}, pp.~189--194.

\bibitem{DecanMMG22}
{\sc Decan, A., Mens, T., Mazrae, P.~R., and Golzadeh, M.}
\newblock On the use of github actions in software development repositories.
\newblock In {\em {IEEE} International Conference on Software Maintenance and Evolution, {ICSME} 2022, Limassol, Cyprus, October 3-7, 2022\/} (2022), {IEEE}, pp.~235--245.

\bibitem{ademeusages2021}
{\sc ENERTECH, ADEME, R.}
\newblock {PANEL USAGES ELECTRODOMESTIQUES - Consommations électrodomestiques françaises basées sur des mesures collectées en continu dans 100 logements}.
\newblock Tech. rep., 03 2021.

\bibitem{ErichAD17}
{\sc Erich, F., Amrit, C., and Daneva, M.}
\newblock A qualitative study of devops usage in practice.
\newblock {\em Journal of Software: Evolution and Process 29}, 6 (2017).

\bibitem{FieniRS20}
{\sc Fieni, G., Rouvoy, R., and Seinturier, L.}
\newblock Smartwatts: Self-calibrating software-defined power meter for containers.
\newblock In {\em 20th {IEEE/ACM} International Symposium on Cluster, Cloud and Internet Computing, ({CCGrid})\/} (Melbourne, Australia, May 2020), {IEEE}, pp.~479--488.

\bibitem{GolzadehDM22}
{\sc Golzadeh, M., Decan, A., and Mens, T.}
\newblock On the rise and fall of {CI} services in github.
\newblock In {\em 29th {IEEE} International Conference on Software Analysis, Evolution and Reengineering ({SANER})\/} (Honolulu, {HI}, {USA}, March 2022), {IEEE}, pp.~662--672.

\bibitem{guennebaud24}
{\sc Guennebaud, G., and Bugeau, A.}
\newblock {Energy consumption of data transfer: Intensity indicators versus absolute estimates}.
\newblock {\em {Journal of Industrial Ecology}\/} (June 2024).

\bibitem{HiltonTHMD16}
{\sc Hilton, M., Tunnell, T., Huang, K., Marinov, D., and Dig, D.}
\newblock Usage, costs, and benefits of continuous integration in open-source projects.
\newblock In {\em 31st {IEEE/ACM} International Conference on Automated Software Engineering, {ASE}\/} (Singapore, September 2016), {ACM}, pp.~426--437.

\bibitem{humble2018accelerate}
{\sc Humble, J., and Kim, G.}
\newblock {\em Accelerate: The science of lean software and devops: Building and scaling high performing technology organizations}.
\newblock IT Revolution, 2018.

\bibitem{JayOLTOF23}
{\sc Jay, M., Ostapenco, V., Lef{\`{e}}vre, L., Trystram, D., Orgerie, A., and Fichel, B.}
\newblock An experimental comparison of software-based power meters: focus on {CPU} and {GPU}.
\newblock In {\em 23rd {IEEE/ACM} International Symposium on Cluster, Cloud and Internet Computing ({CCGrid})\/} (Bangalore, India, May 2023), {IEEE}, pp.~106--118.

\bibitem{kamiya2020data}
{\sc Kamiya, G.}
\newblock Data centres and data transmission networks, 2020.

\bibitem{kinsman2021software}
{\sc Kinsman, T., Wessel, M.~S., Gerosa, M.~A., and Treude, C.}
\newblock How do software developers use github actions to automate their workflows?
\newblock In {\em 18th {IEEE/ACM} International Conference on Mining Software Repositories ({MSR})\/} (Madrid, Spain, May 2021), {IEEE}, pp.~420--431.

\bibitem{wakatareKKSHIK19}
{\sc Lwakatare, L.~E., Kilamo, T., Karvonen, T., Sauvola, T., Heikkil{\"{a}}, V., Itkonen, J., Kuvaja, P., Mikkonen, T., Oivo, M., and Lassenius, C.}
\newblock Devops in practice: {A} multiple case study of five companies.
\newblock {\em Information and Software Technology 114\/} (2019), 217--230.

\bibitem{malladiNPLKH12}
{\sc Malladi, K.~T., Nothaft, F.~A., Periyathambi, K., Lee, B.~C., Kozyrakis, C., and Horowitz, M.}
\newblock Towards energy-proportional datacenter memory with mobile {DRAM}.
\newblock In {\em 39th International Symposium on Computer Architecture {(ISCA} 2012), June 9-13, 2012, Portland, OR, {USA}\/} (2012), {IEEE} Computer Society, pp.~37--48.

\bibitem{malmodin2018energy}
{\sc Malmodin, J., and Lund{\'e}n, D.}
\newblock The energy and carbon footprint of the global ict and e\&m sectors 2010--2015.
\newblock {\em Sustainability 10}, 9 (2018), 3027.

\bibitem{manotas_empirical_2016}
{\sc Manotas, I., Bird, C., Zhang, R., Shepherd, D., Jaspan, C., Sadowski, C., Pollock, L., and Clause, J.}
\newblock An empirical study of practitioners' perspectives on green software engineering.
\newblock In {\em 38th {IEEE/ACM} International Conference on Software Engineering ({ICSE})\/} (New York, {NY}, {USA}, 2016), {ICSE} '16, {ACM}, pp.~237--248.

\bibitem{mcqueen2020bike}
{\sc McQueen, M., MacArthur, J., and Cherry, C.}
\newblock The e-bike potential: Estimating regional e-bike impacts on greenhouse gas emissions.
\newblock {\em Transportation Research Part D: Transport and Environment 87\/} (2020), 102482.

\bibitem{Rub1}
{\sc Menage, P.}
\newblock Control groups, 2023.
\newblock Accessed: 29-07-2023.

\bibitem{emissionsWorld}
{\sc of~Energy, I.~A.}
\newblock Global energy \& co2 status report 2019 - emissions, 2019.
\newblock Accessed: 04-12-2023.

\bibitem{prieto2022energy}
{\sc Prieto, B., Escobar, J.~J., G{\'o}mez-L{\'o}pez, J.~C., D{\'\i}az, A.~F., and Lampert, T.}
\newblock Energy efficiency of personal computers: a comparative analysis.
\newblock {\em Sustainability 14}, 19 (2022), 12829.

\bibitem{owid-travel-carbon-footprint}
{\sc Ritchie, H.}
\newblock Which form of transport has the smallest carbon footprint?
\newblock {\em Our World in Data\/} (2023).

\bibitem{VasilescuYWDF15}
{\sc Vasilescu, B., Yu, Y., Wang, H., Devanbu, P.~T., and Filkov, V.}
\newblock Quality and productivity outcomes relating to continuous integration in github.
\newblock In {\em 10th Joint Meeting on Foundations of Software Engineering, {ESEC/FSE}\/} (Bergamo, Italy, August - September 2015), {ACM}, pp.~805--816.

\bibitem{VassalloPGP19}
{\sc Vassallo, C., Proksch, S., Gall, H.~C., and Penta, M.~D.}
\newblock Automated reporting of anti-patterns and decay in continuous integration.
\newblock In {\em 41st {ACM/IEEE} International Conference on Software Engineering, {ICSE}\/} (Montreal, QC, Canada, May 2019), {IEEE} / {ACM}, pp.~105--115.

\bibitem{VassalloPJGP20}
{\sc Vassallo, C., Proksch, S., Jancso, A., Gall, H.~C., and Penta, M.~D.}
\newblock Configuration smells in continuous delivery pipelines: a linter and a six-month study on gitlab.
\newblock In {\em 28th {ACM} Joint European Software Engineering Conference and Symposium on the Foundations of Software Engineering {ESEC/FSE}\/} (Virtual, November 2020), {ACM}, pp.~327--337.

\bibitem{Zaidman24}
{\sc Zaidman, A.}
\newblock An inconvenient truth in software engineering? the environmental impact of testing open source java projects.
\newblock In {\em 5th {ACM/IEEE} International Conference on Automation of Software Test ({AST})\/} (Lisbon, Portugal, April 2024), {ACM}, pp.~214--218.

\bibitem{ZampettiVPCGP20}
{\sc Zampetti, F., Vassallo, C., Panichella, S., Canfora, G., Gall, H.~C., and Penta, M.~D.}
\newblock An empirical characterization of bad practices in continuous integration.
\newblock {\em Empirical Software Engineering 25}, 2 (2020), 1095--1135.

\end{thebibliography}

\end{document}